\documentclass[twocolumn,amsmath,amssymb,aps]{revtex4}

\usepackage{times}
\usepackage{amsmath}
\usepackage{amssymb}
\usepackage[]{graphicx}
\usepackage{bm}
\usepackage{color}
\usepackage{soul}
\usepackage{float}
\usepackage{natbib}
\begin{document}

\title{Inertial Range Scaling in Rotations of Long Rods in Turbulence }

\author{Shima Parsa, Greg A. Voth$^\ast$\\
Department of Physics, Wesleyan University,\\ Middletown, Connecticut 06459, USA\\
$^\ast$ Correspondence to: gvoth@wesleyan.edu
}

\date{\today}

\begin{abstract}

We measure the rotational statistics of neutrally buoyant rods with lengths $2.8 < l/\eta <72.9 $ in turbulence.  For particles with length in the inertial range,  we derive a scaling relationship for the mean square rotation rate, $\langle \dot{p}_i \dot{p}_i \rangle \propto l^{-4/3}$ and show that measurements approach this scaling.  Deviations from the proposed scaling are explained as the effect of dissipation range scales. The correlation time of the Lagrangian autocorrelation of rod rotation rate scales as the turn over time of eddies of the size of the rod.  Measuring rotational dynamics of single long rods provides a new way to access the spatial structure of the flow at different length scales.
\end{abstract}
\maketitle
The dynamics of particulate material in fluid flows is important in a broad range of problems in nature \cite{Bowen1993} and industry \cite{Lundell2011}. Particles that are small and neutrally buoyant exhibit the same translational dynamics as fluid elements and are considered Lagrangian tracers. However, in many cases the particulate matter has different dynamics from fluid elements because it is not density matched with the fluid or has sizes  larger than the smallest length scales of the flow. The motion of density mismatched spherical particles in fluid flows has received a lot of attention~\cite{Falkovich2002,Shaw2003,Toschi2009,Meiburg2010}.     Recently, the motion of neutrally buoyant, large, spherical particles in turbulent flows have become accessible in experiments \cite{voth2002a, Qureshi2007, Brown2009, Volk2011} and numerical simulations~\cite{Homann2010, Calzavarini2012}, and this has provided a novel method by which the spatial structure of turbulent flows can be probed with single particle measurements.   The dynamics of spheres that are both large and density mismatched is more complex and has been addressed by some very recent experiments~\cite{Prakash2012}.

Anisotropic particles have very different translational and rotational dynamics from spherical particles.  Measurements and simulations of small rod-like particles show preferential alignment with the velocity gradients of the flow~\cite{Pumir2011,Parsa2011} and this alignment suppresses the measured rotation rate in turbulent flows~\cite{shin2005,Parsa2012}.  A wide range of experimental and numerical studies have explored the dynamics of neutrally buoyant small rods and fibers in different flows~\cite{Parsheh2005, Holm2007, Lin2007,Mortensen2008b,Marchioli2010}.  Only a few studies have focused on the dynamics of long rods in turbulence, where rod length spans over many times $\eta$, the Kolmogorov length scale~\cite{olson1998,shin2005}. Using numerical simulations and slender body theory, Shin and Koch \cite{shin2005} studied the translational and rotational dynamics of long fibers in turbulent flow at Taylor Reynolds number up to $R_\lambda=53.3$.   Among other things they show how the mean square rotation rate decreases as rods become longer than the tracer limit and identify the key role played by alignment of rods.

It is known that the acceleration variance of large spheres scales with their diameter approximately as $\langle a^2 \rangle \sim d^{-2/3}$ for \emph{d} in the inertial range \cite{voth2002a, Qureshi2007, Brown2009, Volk2011}.   This result can be obtained by dimensional arguments simply assuming that the sphere terminates the cascade at its diameter~\cite{voth2002a}.   It can also be derived from the inertial range form of the pressure structure function or the acceleration correlations~\cite{Qureshi2007,Brown2009,Volk2011}.  This suggests that measuring the acceleration of large spheres provides access to the statistics of the turbulence at scales equal to the size of the spheres using only single particle measurements.   Recent experimental work by Volk et al~\cite{Volk2011} finds that the $d^{-2/3}$ scaling is not exact and proposes an improved scaling of $d^{-0.81}$ by including intermittency in the pressure structure function.  The fact that the accelerations of large spheres provides information about inertial range scaling suggests that similar information should be available in the rotations of long rods.

In contrast to the point-particle models used for small inertial particles, the equations of motion for large spheres are largely unknown.  Fax\'{e}n corrections can be used to extend point particle models to describe large spheres~\cite{Calzavarini2009}, but these models have difficulties when particles are much larger than $\eta$~\cite{Homann2010}.

For rods, the analytical work can be done more rigorously to connect the motion of rods to the fluid motion. Olson and Kerekes~\cite{olson1998} have introduced a model to describe the rotational velocity of fibers. This model is based on the assumption that the fibers are infinitely thin and composed of a number of sections smaller than $\eta$ which are hydrodynamically independent.  For a neutrally buoyant fiber of length \emph{l}, the rotation rate is
\begin{equation}
\dot{p}_i = \frac{12}{l^3} \int_{-l/2}^{l/2}\left(\delta_{ij}-p_ip_j\right)u_j(r)r dr
\label{Eq:rotLong}
\end{equation}
where \textbf{p} is the orientation unit vector of the fiber and \textbf{u} is the turbulent velocity at points along the fiber. If the orientation of a rod is uncorrelated with the velocity field in Eq.~\ref{Eq:rotLong}, then the mean square rotation rate of randomly oriented long rods is~\cite{olson1998}
\begin{equation}
\langle\dot{p_i}\dot{p_i}\rangle = \frac{48 \tilde{u}^2}{l^3} \int_{0}^{l} \left[1- 3\frac{r}{l}+2 \left(\frac{r}{l}\right) ^3 \right]R_{NN}(r) dr
\label{Eq:rotRandom}
\end{equation}
where $\tilde{u}^2$ is the rms velocity of the fluid flow and $R_{NN}(s)$ is the fluid transverse velocity correlation function at separation distance of~\emph{r}. Shin and Koch~\cite{shin2005} show that Eq.~\ref{Eq:rotRandom} is in good agreement with their simulations for the case of randomly oriented rods.

\begin{center}
\begin{table*}
\begin{tabular*}{0.8\textwidth}
     {@{\extracolsep{\fill}}|c|c|c|c|c|c|}
\hline
$\bm{R_\lambda}$&$\bm{\tilde{u}}$&&$\bm{L}$&$\bm{\eta}$&$\bm{\tau_\eta}$\\
$=(15 \tilde{u} L/\nu)^{1/2}$&$\sqrt{u_i u_i/3}$&\raisebox{1.5ex}{$\bm{\langle\epsilon\rangle}$}&$=\tilde{u}^3/\langle\epsilon\rangle$&$=(\nu^3 / \langle\epsilon\rangle)^{1/4}$&$=(\nu/\langle\epsilon\rangle)^{1/2}$\\

\hline
&mm/s&mm$^2$/s$^3$&mm&mm&s\\
\hline
150&30.4&319&87.9&0.36&0.074\\
210&62.8&2800&84&0.21&0.025\\
\hline
\end{tabular*}
\caption{ Table of flow parameters: $R_\lambda$, Taylor Reynolds number; $\tilde{u}$, rms velocity of the flow; $L$, energy input length scale; $\langle\epsilon\rangle$, energy dissipation rate; $L$, energy input length scale; $\eta$, Kolmogorov length scale; $\tau_\eta$, Kolmogorov time scale. $\nu$ is the fluid kinematic viscosity and is 1.75$\times$10$^{-6}$ m$^2$/s.}
\label{tabel1}
\end{table*}
\end{center}

In this paper, we introduce a scaling for the mean square rotation rate of rods with lengths in the Kolmogorov inertial range.   This scaling can be obtained either from dimensional arguments or from Eq.~\ref{Eq:rotRandom}.   The rotation rate has dimensions of inverse time, so for tracer rods it scales with the Kolmogorov time scale, $\langle \dot{p_i}\dot{p_i}\rangle\sim{\tau_\eta}^{-2}$. Assuming that long rods are only rotated and aligned by eddies close to their size, the mean square rotation rate for rods at length scale \emph{l} will scale like ${\tau_l}^{-2}$, where $\tau_l$ is the time scale of eddies of size \emph{l}. In the inertial range the time scale $\tau_l$ can be defined as $\tau_l= l/u_l =l/(l\langle\epsilon\rangle)^{1/3})$, where $u_l$ is the velocity at length \emph{l}, and $\langle \epsilon \rangle$ is the mean energy dissipation rate.
This dimensional argument gives $\langle \dot{p_i}\dot{p_i}\rangle\sim l^{-4/3}$ for $l$ in the inertial range.

The same $l^{-4/3}$ scaling can be obtained using Eq.~\ref{Eq:rotRandom}, with the additional benefit that the coefficient of the scaling law can be determined in terms of the coefficient of the second order structure function.    In homogeneous turbulence the transverse correlation function is given in terms of the transverse velocity structure function,  $\tilde{u}^2R_{NN}(r) = \tilde{u}^2-1/2 D_{NN}(r)$.  In the inertial range of isotropic turbulence,  $D_{NN}(r)=  (4/3) C_2 (\langle \epsilon\rangle r)^{(2/3)}$, where $C_2$ is an approximately universal constant \cite{Pope, Sreenivasan1995}. Using this form in Eq.~\ref{Eq:rotRandom}, the mean squared rotation rate for randomly oriented rods in the inertial range is
\begin{equation}
\frac{\langle \dot{p_i}\dot{p_i}\rangle }{ (\langle \epsilon \rangle/\nu)} =\frac{108}{35} C_2 \left(\frac{l}{\eta}\right)^{-4/3}
\label{eq:scaling_constant}
\end{equation}
where $\nu$ is the kinematic viscosity.  Here we use the fact that only scales near the length of the rod contribute to the rotation rate in Eq. \ref{Eq:rotRandom} so that in the high Reynolds number limit the inertial range form of the structure function can be used for all $r$.

We have performed  a series of experiments to measure the rotation of rods in 3D turbulence for rod lengths that extend from the dissipation range well into the inertial range to explore whether an $l^{-4/3}$ scaling range exists.  We measure neutrally buoyant rod like particles with lengths ranging from $2.8 \eta$ up to $72.9\eta$ in a turbulent flow between oscillating grids \cite{Blum2010}. The rods are nylon thread  with diameter of 0.2 mm and are cut to different lengths (\emph{l} =1, 3, 6.8, 15.2 mm). All particles are dyed fluorescent for better detection.  The rotational dynamics of rods are measured using stereoscopic images from four high speed cameras\cite{Parsa2012}.  Each camera that detects a rod defines a plane in which the rod should exist. We require at least three cameras to detect each single rod and the orientation of the rod is the intersection of the planes these cameras define.   The detection volume is illumined with 4 laser beams.  This has nearly removed a limitation in earlier experiments~\cite{Parsa2012} where the probability of detecting a particle depended on the orientation of the particle with respect to the laser beam.

\begin{figure}[b]
 \centerline{\includegraphics[scale=.38]{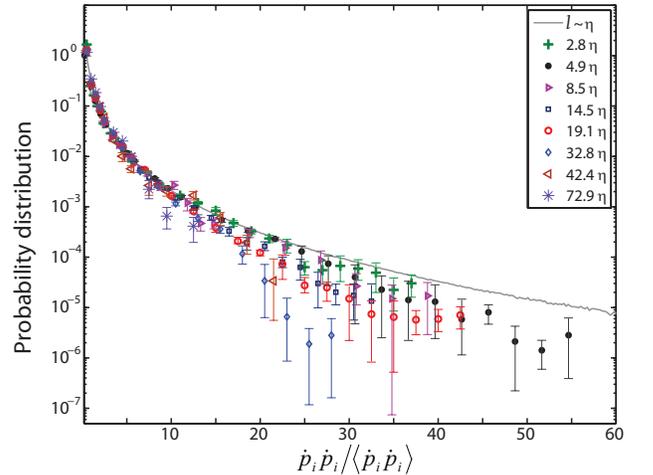}}
\caption{ (Color online) The PDF of rotation rate squared for different rod lengths. The lengths of rods are $l/\eta$= 2.8 (green crosses), 4.9 (black filled circles), 8.5 (right triangles), 14.5 (open squares), 19.1 (red open circles), 32.8 (blue diamond), 42.2 (brown left triangles) and 72.9 (purple asterisk) and tracers from simulation (solid gray line). The results are reported from two experiments at $R_\lambda=150$ and 210. The simulations\cite{Parsa2012} are for tracer rods at $R_\lambda$=180.
}
\label{Fig:PDFP2long}
\end{figure}

The rotation rate vector of rods, $\dot{\textbf{p}}$, is measured from quadratic fits to the measured orientations along trajectories versus time.  The measurements are at two different Taylor Reynolds numbers ($R_\lambda$ =150, 210).   We have done parallel experiments with tracer particles to measure the turbulence parameters. From measured tracer velocities, we extract the second and third order longitudinal structure functions and obtain the energy dissipation rate, $\langle\epsilon\rangle$, from Kolmogorov's 4/5 law. We obtain the flow parameters shown in Table~\ref{tabel1} from the measured energy dissipation rate and rms flow velocity.

In these experiments the maximum detection volume is 160 cm$^3$, while the effective detection volume is different for each rod length and is the smallest (85 cm$^3$) for the longest rods at 15.2 mm. The effective detection volume is smaller to ensure that the entire rod is in the illuminated detection volume so the position and orientation of rods are measured more accurately based on the full length of rods. The number density of rods is small so particle-particle interaction is negligible. The particle concentration is 0.025~cm$^{-3}$ for 1 mm rods, and 0.0075~cm$^{-3}$ for the longest rods at 15.2 mm. The uncertainty in measuring the center of rods is 60$\mu$m for the 1 mm rods and 180 $\mu$m for 15.2 mm rods.  This uncertainty is determined based on the stereomatching accuracy. The accuracy of measuring the orientation for 1 mm particles is 0.01 rad and increases with rod length. The uncertainty in the orientation of the rods is determined from the residual of the intersection of planes defined by multiple cameras.

Figure~\ref{Fig:PDFP2long} shows the probability distribution function (PDF) of the rotation rate squared, $\dot{p}_i\dot{p}_i$, normalized by the mean for different rod lengths. The PDF shows only a weak dependence on the rod length and all rod lengths show rare events with large rotation rates.  The probability of rare events is somewhat smaller for long rods ($l/\eta>20$) than for tracer rods ($l/\eta<7$); however, this difference is slightly larger than the measurement uncertainty due to the smaller number of samples for long rods. The error bars represent the random statistical error and the systematic error in measuring the rotation rates.  Qualitatively, the rotation rate PDF depends on rod length in the same way that the acceleration PDF for large spheres depends on diameter~\cite{Volk2011},  with only a small narrowing of the tails for large particles.

\begin{figure}[tb]
 \centerline{\includegraphics[scale=.37]{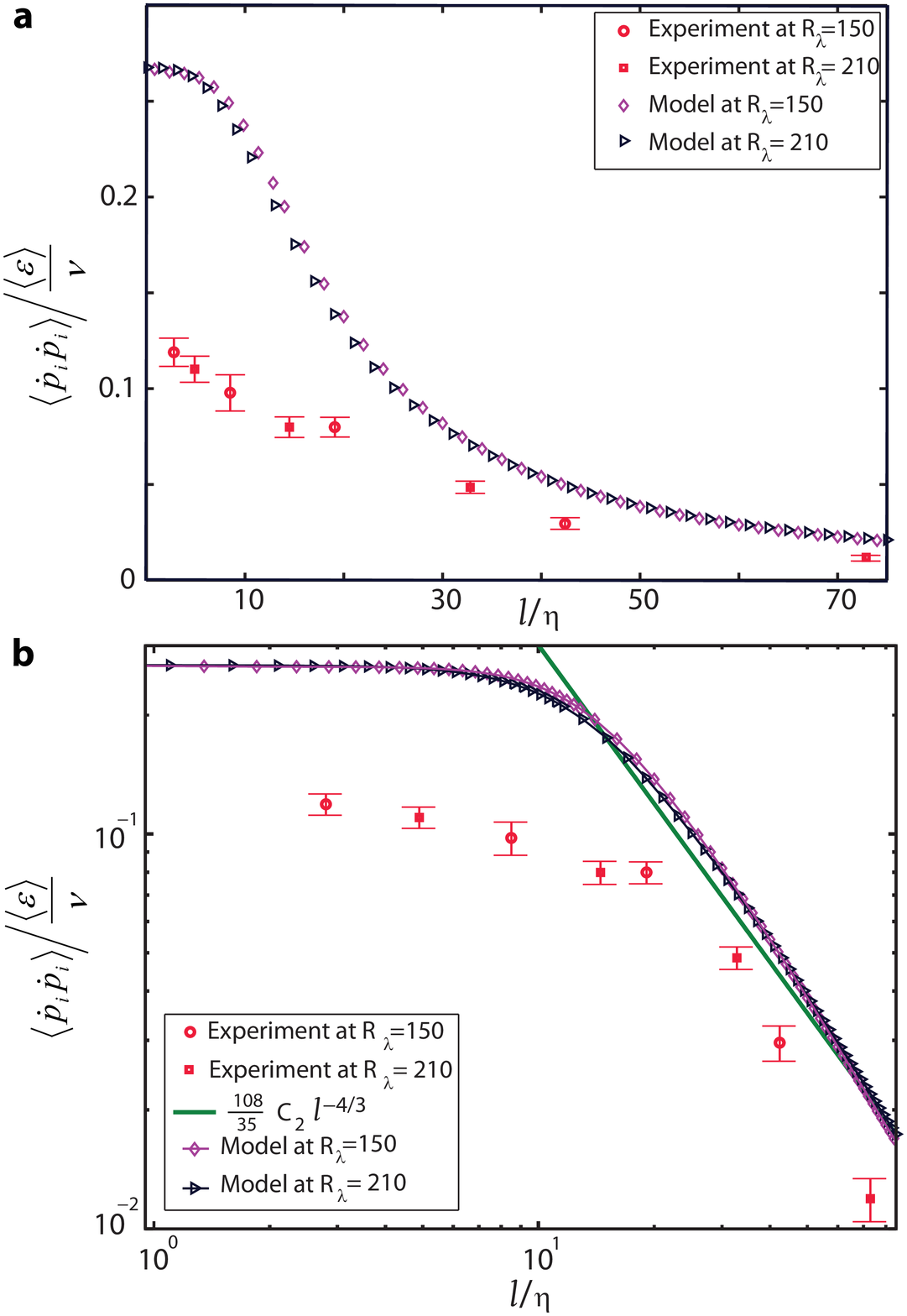}}
\caption{ (Color online) Mean square rotation rate as a function of rod length. a) Comparison of the experimental data at $R_\lambda$ =150 (red open circles) and 210 (red squares) with the model of randomly oriented rods at $R_\lambda$=150 (purple diamond) and $R_\lambda$= 210 (black triangles).  b) Comparison of the mean square rotation rate of rods (red open circles and squares) and the model for randomly oriented rods (diamond and triangles) with an $l^{-4/3}$ inertial range scaling law (solid green line).
}
\label{Fig:variancep2}
\end{figure}

The mean square rotation rate for different rod lengths is shown in Fig.~\ref{Fig:variancep2}(a). Our experimental measurements show that by increasing the length of rods, the mean square rotation rate decreases.   This is expected since longer rods should begin to filter out the contributions from some eddies smaller than their length.   These results agree with an earlier simulation\cite{shin2005} of long fibers in turbulent flow at $R_\lambda$ =53.3.   The experimental measurements are slightly larger, but the difference is only a little larger than the measurement uncertainty.

For a more detailed understanding of the dependence of the rotation rate on the length of rods, we compare these results with the rotation rate of randomly oriented rods. Assuming the orientation of rods is uncorrelated with the fluid velocity, Eq.~\ref{Eq:rotRandom} can be used to calculate the mean square rotation rate from measured second order structure functions.  Figure~\ref{Fig:variancep2}(a) shows the mean square rotation rate for randomly oriented rods predicted by this model using our experimental measurements of $D_{NN}$ at $R_\lambda $=150, 210.
The mean square rotation rate of short rods is much smaller than randomly oriented rods of the same length ($l/\eta$). However, this difference decreases as the length of the rod is increased. Previous studies of tracer rods~\cite{girimaji1990,Luthi2005,shin2005,Pumir2011,Parsa2012} have shown that as rods are carried by the flow their orientation becomes correlated with the directions defined by the velocity gradient tensor of the flow.  This alignment results in suppression of the rotation rate of short rods compared to randomly oriented rods \cite{Parsa2012}. The smaller differences between the measured rotation rates of long rods and randomly oriented rods suggests that the alignment is weaker between long rods and the velocity gradients responsible for rotating the long rods.   In the previous simulations of long fibers \cite{shin2005} the correlation of long fibers with the flow is much weaker than what we see in the experiment, possibly as a result of their much smaller Reynolds number.

Figure~\ref{Fig:variancep2}(b) shows the inertial range scaling law from Eq.~\ref{eq:scaling_constant} with $C_2=2.0$ and  compares it with the experimentally measured rotation rates and the prediction of Eq.~\ref{Eq:rotRandom} for randomly oriented rods using the measured velocity structure function.   Both the measured rotation rates and the prediction of Eq.~\ref{Eq:rotRandom} approach an $l^{-4/3}$ scaling for large $l$.  The experimental data has a smaller coefficient compared to randomly oriented rods as expected due to alignment effects.  Within error bars of the experimental data for $l > 30 \eta$, one could also fit the data with a different exponent slightly steeper than  $-4/3$.  In this same range , the prediction of Eq.~\ref{Eq:rotRandom} is also steeper than $l^{-4/3}$.  The cause of the steeper scaling can be found in the fact that the prediction of Eq.~\ref{Eq:rotRandom} overshoots the power law scaling in the range $20\eta <l <50\eta$.  This overshoot occurs in the range of scales slightly larger than the dissipative range because scales smaller than the length of the rod contribute to the rotation rate.    For $l> 50 \eta$, well into the inertial range, the prediction converges with the scaling law from Eq.~\ref{eq:scaling_constant} because here the contributions from the dissipation range are becoming negligible.    The overshoot also appears to exist in the experimental rotation rate data, although the effect is of the same size as the measurement uncertainties.

The slightly steeper scaling of the rod rotation rate is similar to the effect observed in Ref.~\cite{Volk2011} for the accelerations of spheres with diameters in the inertial range.  They argue that intermittency effects are responsible for the difference between their measured scaling and the prediction of dimensional analysis.  It is possible that intermittency effects also play a role for rods.   However, the availability of a solid theoretical foundation for calculating rotation rates of long rods suggests another possible explanation due to effects of small scales.   We have used the Batchelor parameterization~\cite{Batchelor1951,Grossmann1995} of the structure function at very high Reynolds number in Eq.~\ref{Eq:rotRandom} and find an overshoot which leads to steeper scaling for  $30 \eta < l < 100 \eta$ and then an agreement with $l^{-4/3}$ for $l > 100 \eta$.  This parameterization has no intermittency effects.   It is possible that spheres also have an overshoot in which particles only a little larger than the dissipation scale have accelerations larger than the inertial range power law due to the contributions from dissipation range scales.

\begin{figure}[tb]
 \centerline{\includegraphics[scale=.37]{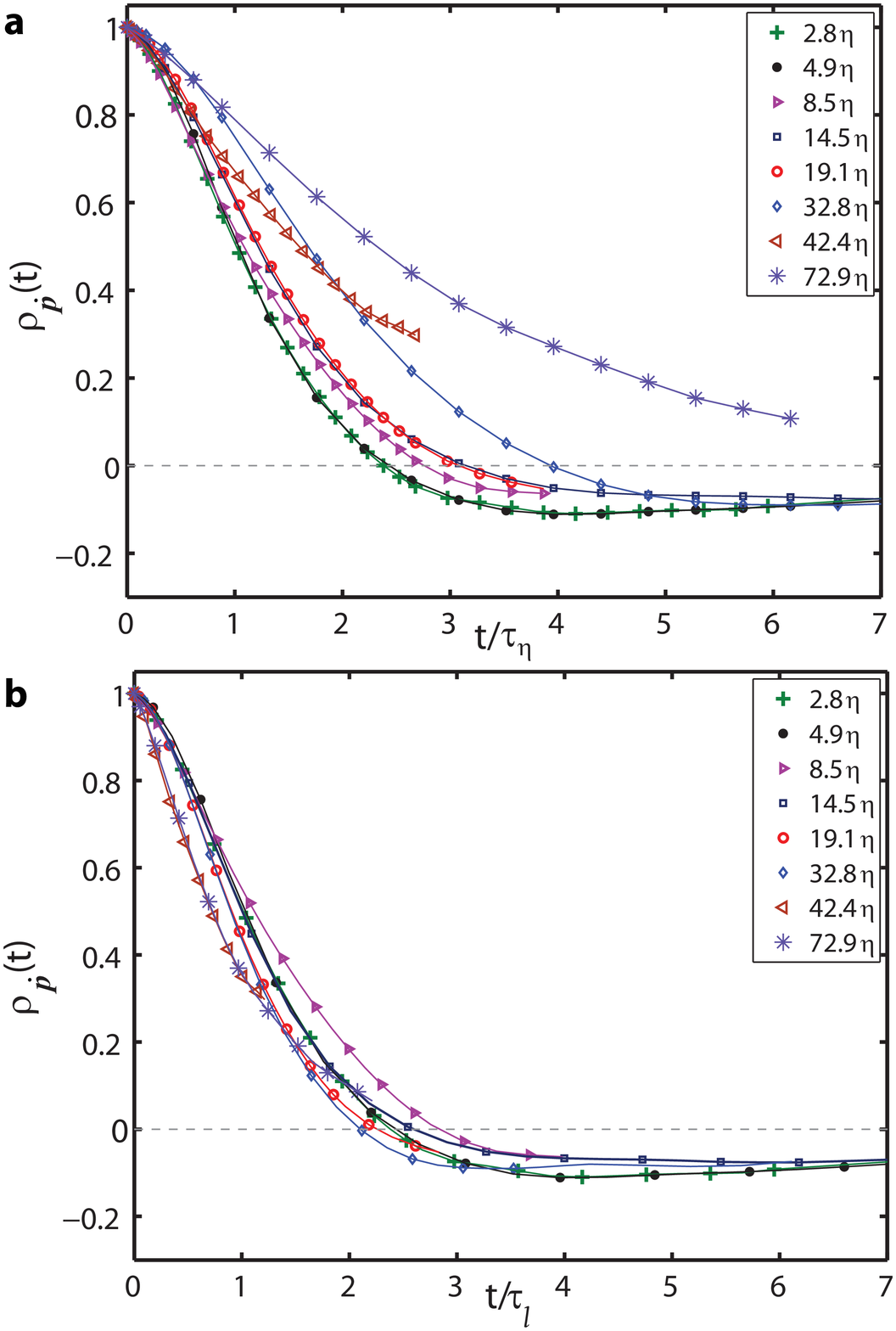}}
\caption{(Color online) Lagrangian autocorrelation of rotation rate for different rod length.  The lengths of rods are $l/\eta$= 2.8 (green crosses), 4.9 (black filled circles), 8.5 (right triangles), 14.5 ( open squares), 19.1 (red open circles), 32.8 (blue diamond), 42.2 (brown left triangles) and 72.9 (purple asterisk). a) Time is normalized by the Kolmogorov time.  b) Time is normalized by the turn over time of eddies with size equal to the length of the rods, $\tau_\textit{l}$. The symbols are displayed at every other data point.
}
\label{Fig:RotationAuto}
\end{figure}
To measure the mean square rotation rate, we use the extrapolation method developed in Ref.~\cite{voth2002a} to remove the dependence on the smoothing used to measure derivatives. The extrapolation to zero smoothing length is known to overestimate the actual rotation rate, and we performed a simulation of our measurement system to estimate the size of this overestimate.  The translational motion of tracer rods matches that of fluid particles, so we are able to use a database of previously simulated Lagrangian trajectories \cite{Benzi2009} to integrate Jeffery's equation \cite{Jeffery1922} and obtain the orientation of tracer rods along Lagrangian trajectories. We have used these trajectories to create simulated experimental images of rods and analyzed these images with the same analysis codes used for the experimental data. The extrapolated mean square rotation rate of simulated rods is found to be larger than direct measurements of the DNS rods by 6\%.  We have corrected for this by shifting our mean square rotation rate measurements down by 6\%.
The error bars on the experimental measurement of the mean square rotation rate in Fig. \ref{Fig:variancep2} represent both statistical uncertainty and the systematic error due to the overestimation correction.

The small difference between two Reynolds numbers in the model for the mean square rotation rate of randomly oriented rods in Fig. \ref{Fig:variancep2} is due to uncertainties in measuring the second order transverse velocity structure function. As there are very few tracer particle pairs measured with very small separations ($r < 15 \eta$), we used analytical forms of the structure function there.
For $r < \eta$, the dissipation range is $D_{NN}(r)= \frac{2}{15} (\langle\epsilon\rangle/\nu)r^2$. For $\eta < r<15\eta$, we use a fit of the Batchelor parametrization \cite{ Batchelor1951,Grossmann1995}. This method provides a smooth transition from small scales to the region of experimentally resolved velocity structure function.

Figure~\ref{Fig:RotationAuto}(a) shows the Lagrangian autocorrelation of rotation rate measured for different rod lengths. Our measurements show that the correlation time of the rotation rate depends on the length of rods and increases with rod length.  Similar autocorrelation functions were obtained in simulations~\cite{shin2005} at $R_\lambda$=39.9.   We expect that if the rods are rotating due to eddies of their size, \emph{l}, then the  decay time for the correlation of rotation rate should scale as the turn over time at the length of the rods. In Fig.~\ref{Fig:RotationAuto}(b) the horizontal axis (\emph{t}) is normalized by $\tau_l$, time scale of eddies with length scale \emph{l}. After this normalization, the Lagrangian autocorrelation of rotation rate for all rod lengths collapse on a single curve within measurement uncertainty. The time scales for short tracer rods ($l/\eta<5$) is the Kolmogorov time-scale ($\tau_\eta$) and the time scales for longer rods are measured from longitudinal second order velocity structure function of  the fluid particles ($\tau_l= \frac {1}{\sqrt{15}} l/\delta u_l =\frac {1}{\sqrt{15}} l/\sqrt{D_{LL}(l)}$). Measuring the autocorrelation function for long rods ($l=42.4 \eta$ and $72.9 \eta$) is difficult as the effective detection volume is small compared to the length of these rods so the trajectories are not long enough to measure long time autocorrelations.   The collapse seen in Fig.~\ref{Fig:RotationAuto}(b)  provides additional evidence, beyond that seen in the mean square rotation rate data,  that rod rotations are controlled by eddies with size near the rod length.

Long rods provide a promising path for studying the dynamics of large particles in turbulence.   For rods, slender body theory can be used to connect particle motion with the fluid flow even for particles much larger than the Kolmogorov scale while the analytical results for spheres are only available for small deviations from the small particle limit.   We have  presented measurements of the rotations of rods with lengths extending well into the inertial range, and we find that the mean square rotation rate in the inertial range approaches the $l^{-4/3}$ scaling that is expected from inertial range scaling of the velocity structure functions.   The PDF of rotation rate shows only a weak dependence on rod length.   We find that rods develop preferential alignment so that their rotation rates are significantly smaller than that predicted for randomly oriented rods.  This alignment depends on rod length as rods in the inertial range show a smaller effect of alignment than tracer rods.  The Lagrangian autocorrelation time of the rotation rate depends on the length of rods and scales with the eddy turn over time at a scale equal to the rod length.

Shin and Koch \cite{shin2005} provided a groundbreaking simulation data set on this problem, but were limited to $R_\lambda < 53.3$ where there is essentially no inertial range.  Future high Reynolds number DNS using their method of simulating long fibers  offers the possibility
to study the motion of long rods while also having access to the full velocity field around the rods.  Experimental tracking of long rods in turbulence allows access to the dynamics of turbulent scales at the length of the particle from single particle measurements, and has potential to provide valuable information about Lagrangian dynamics as a function of scale in complex turbulent flows.

\noindent We acknowledge support from  NSF grant DMR-1208990, and COST Actions MP0806 and FP1005. We thank Stefan Kramel for assistance in the experimental work and Enrico Calzavarani, Federico Toschi, Nicholas T. Ouellette and Rui Ni for stimulating discussions.


\end{document}